\documentclass[twoside,twocolumn,9pt,amsmath,amssymb]{article}
\usepackage{graphicx} 
\usepackage[utf8x]{inputenc}
\usepackage[T1]{fontenc}
\usepackage{verbatim} 
\usepackage{amsmath} 
\usepackage{xcolor}

\begin{document}

\title{Gas pressure in bubble attached to tube circular outlet}

\author{A. Salonen$^{1}$ \and C. Gay$^{2}$ \and A. Maestro$^{1,3}$ \and W. Drenckhan$^{1}$ \and E. Rio$^{1}$}
\date{\today}
\maketitle

       {
         $^{1}$ Universit\'e Paris-Sud, Laboratoire de Physique des Solides, UMR8502, Orsay, F-91405
         
  $^{2}$ Universit\'e Paris~Diderot--Paris~7
Mati\`ere et Syst\`emes Complexes (CNRS UMR 7057),
B\^atiment Condorcet, Case courrier 7056, 75205 Paris Cedex 13

$^{3}$ Present address: Biological and Soft Systems, University of Cambridge, UK.
}\newline\newline

\newcommand{\hs}{\hspace{0.7cm}}
\newcommand{\be}{\begin{equation}}
\newcommand{\ee}{\end{equation}}
\newcommand{\bee}{\begin{eqnarray}}
\newcommand{\eee}{\end{eqnarray}}
\newcommand{\fin}{\nonumber\\}

\newcommand{\Rapex}{R_{\rm apex}}
\newcommand{\zapex}{z_{\rm apex}}
\newcommand{\rout}{r_{\rm out}}
\newcommand{\zout}{z_{\rm out}}
\newcommand{\sout}{s_{\rm out}}
\newcommand{\pgas}{p_{\rm gas}}
\newcommand{\pliqout}{p_{\rm liq}^{\rm out}}

\newcommand{\rr}{\hat{r}}
\newcommand{\rz}{\hat{z}}
\newcommand{\rs}{\hat{s}}
\newcommand{\rd}{\hat{\delta}}
\newcommand{\rV}{\hat{V}}
\newcommand{\rO}{\hat{\Omega}}
\newcommand{\rzapex}{\hat{z}_{\rm apex}}
\newcommand{\rrout}{\hat{r}_{\rm out}}
\newcommand{\rzout}{\hat{z}_{\rm out}}
\newcommand{\rsout}{\hat{s}_{\rm out}}
\newcommand{\rdout}{\hat{\delta}_{\rm out}}
\newcommand{\ralpha}{\hat{\alpha}}
\newcommand{\rbeta}{\hat{\beta}}

\begin{figure}[h!]
\begin{center}
\resizebox{1.0\columnwidth}{!}{%
\includegraphics{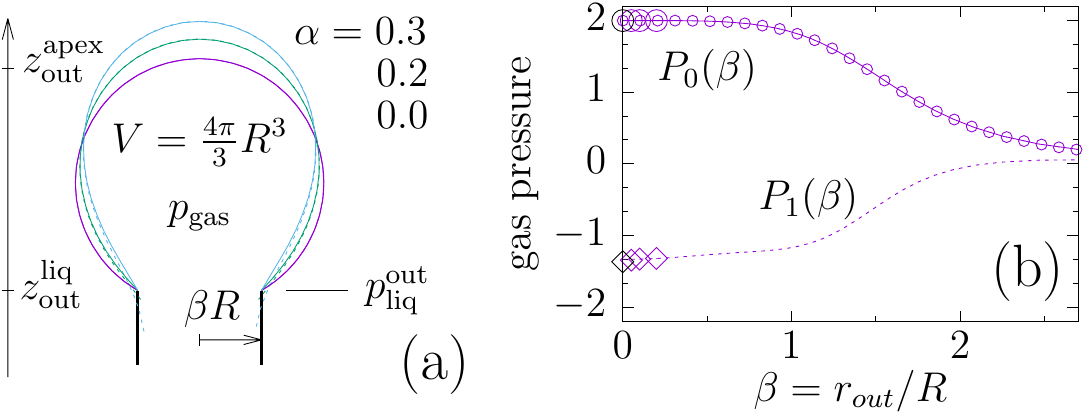}
}
\end{center}
\caption{
  {\em Left:} Numerical simulation of the shape of a bubble
  of volume $4\pi R^3/3$ attached to an outlet of radius $\rout=\beta R$,
  with different intensities of gravity (parameter $\alpha$).
  Exact shapes are plotted as solid lines.
  The shapes obtained at first order in $\alpha$,
  given by Eqs.~(\ref{eq:def:rr1},\ref{eq:def:rz1}),
  are plotted as dashed lines and depart from the exact shapes
  near the outlet.
  {\em Right:} Low gravity limit ($\alpha\rightarrow 0$)
  of the gas pressure $P(\alpha,\beta) \rightarrow P_0(\beta)$
  (upper curve, given by Eq.~\eqref{eq:P0:Om0:1:3}
  and circles for $2A_0$, see Eq.~\eqref{eq:A0:U:beta}) and derivative
  $\partial P/\partial\alpha(\alpha,\beta) \rightarrow P_1(\beta)$
  (lower curve, Eq.~\eqref{eq:dPdalpha:beta})
  as a function of the outlet reduced radius $\beta$.
  Large circles and diamonds represent the values
  obtained numerically in Section~\ref{sec:numeric:shape}.
} 
\label{fig:bubble:a:shape:b:pressure}
\end{figure}

In the present Supplementary notes
to our work ``Arresting bubble coarsening:
A two-bubble experiment to investigate grain growth
in presence of surface elasticity'' (submitted to {\em EPL}),
we derive the expression of the gas pressure $\pgas$
inside a bubble of volume $V$
located above and attached to the outlet
of a tube of radius $\rout$
(see Fig.~\ref{fig:bubble:a:shape:b:pressure}a):
\bee
\pgas &\approx& \pliqout
+ 2 \frac{\gamma}{R} 
- \frac43 \rho g R
\label{eq:pressure:2:frac43}
\eee
where $\pliqout$ is the pressure in the liquid
at the same altitude as the tube outlet
(see Fig.~\ref{fig:bubble:a:shape:b:pressure}a)
and where $R$ is defined from the bubble volume $V$
as the radius of a sphere of volume $V$:
\bee
R &=& \left(\frac{3V}{4\pi}\right)^{1/3}
\eee

\section{Contents of the present notes}

The calculation presented in the present notes
is performed in the limit of low gravity
and is obtained in the following form:
\bee
\pgas &=& \pliqout
+ P_0(\beta) \frac{\gamma}{R} 
+ P_1(\beta) \rho g R
+ ...
\eee
where $\beta$ is the ratio of the outlet to the bubble size:
\bee
\beta &=& \frac{\rout}{R}
\label{eq:def:beta:rout:R}
\eee
More generally, let us define $P(\alpha,\beta)$ through:
\bee
\pgas &=& \pliqout
+ \frac{\gamma}{R} P(\alpha,\beta)
\label{eq:def:nondim:P}
\eee
where $\alpha$ is the non-dimensionalized 
gravity:
\bee
\alpha &=& \frac{\rho g R^2}{\gamma} = \frac{R^2}{l_{\rm cap}^2}
\label{eq:def:alpha:rho:g:R:gamma:lcap}
\eee
Note that for a given value of the outlet radius $r$ (or $\beta$),
there exists a maximum gravity (or $\alpha$)
that must not be exceeded for the bubble
to remain attached to the tube outlet:
\be
0 \leq \alpha < \alpha_{\rm max}(\beta)
\label{eq:alpha:stability:alphamax}
\ee

In the limit of low gravity ($\alpha\rightarrow 0$),
Eq.~\eqref{eq:def:nondim:P} can be expanded as:
\bee
\pgas &=& \pliqout
+ \frac{\gamma}{R} \left(
P_0(\beta) + \alpha P_1(\beta) + {\cal O}(\alpha^2)
\right)
\qquad
\label{eq:P:def:taylor:P0:P1}
\eee
where:
\bee
P_0(\beta) &=& \lim_{\alpha\rightarrow 0}{P(\alpha,\beta)}
\label{eq:def:P0:lim:P}
\\
P_1(\beta) &=& \lim_{\alpha\rightarrow 0}{\frac{\partial P}{\partial\alpha}(\alpha,\beta)}
\label{eq:def:P1:lim:dP:dalpha}
\eee

Functions $P_0$ and $P_1$ are the output of the calculation
performed in the present Supplementary notes.
They are plotted on Fig.~\ref{fig:bubble:a:shape:b:pressure}b.
The small outlet limit ($\beta\rightarrow 0$)
is obtained in Appendix~\ref{sec:appendix:small:outlet:limit}
and is consistent with Fig.~\ref{fig:bubble:a:shape:b:pressure}b:
\bee
P_0(0) &=& 2
\\
P_1(0) &=& -\frac43
\eee
These coefficients are those shown in Eq.~\eqref{eq:pressure:2:frac43}.
Note: this outlet limit ($\beta \rightarrow 0$)
is to be taken {\em after}
the small gravity limit ($\alpha \rightarrow 0$)
which is the basis of the expansion of Eq.~\eqref{eq:P:def:taylor:P0:P1}
and of the whole calculation
of Appendix~\ref{sec:appendix:analytic:shape}.
If the outlet radius goes to zero ($\beta \rightarrow 0$)
before gravity goes to zero, the bubble detaches!

These notes are organized as follows.
In Section~\ref{sec:expression:pressure},
we express the gas pressure in terms of the bubble geometry.
In Section~\ref{sec:bubble:shape},
we calculate the bubble shape equation
and solve it first analytically, then numerically,
and thus obtain the expressions and data
used in Fig.~\ref{fig:bubble:a:shape:b:pressure}
and in Eq.~\eqref{eq:pressure:2:frac43}.

\section{Expression of the gaz pressure}
\label{sec:expression:pressure}

As compared to the pressure $\pliqout$ in the liquid
at the same altitude as the tube outlet,
the gas pressure $\pgas$ can be expressed
for instance in terms of the apex radius of curvature and altitude:
\be
\pgas = \pliqout
+ \frac{2\gamma}{\Rapex} + \rho g (\zout - \zapex)
\label{eq:p:dimension:apex}
\ee
where the last term provides the pressure difference
between the liquid pressure at the tube outlet and apex altitudes
while the middle term expresses the pressure jump
across the interface at the apex,
due to the interfacial tension $\gamma$
and the total curvature $2/\Rapex$.

When the bubble is attached to the tube outlet
($\alpha < \alpha_{\rm max}(\beta)$),
let us define functions $A$ and $B$
through the apex radius of curvature and altitude
that appear in Eq.~\eqref{eq:p:dimension:apex}:
\bee
\Rapex &=& \frac{R}{A(\alpha,\beta)}
\label{eq:apex:r:f1}
\\
\zapex-\zout &=& 2R \, B(\alpha,\beta)
\label{eq:apex:z:f2}
\eee
Substituting Eqs.~(\ref{eq:apex:r:f1},\ref{eq:apex:z:f2})
into Eq.~\eqref{eq:p:dimension:apex}
and combining with Eq.~\eqref{eq:def:nondim:P}:
\be
P(\alpha,\beta) = 2 A(\alpha,\beta) -2 \alpha B(\alpha,\beta)
\ee

Let us expand functions $A$ and $B$ in the limit of low gravity:
\bee
A(\alpha,\beta) &=& A_0(\beta) + A_1(\beta)\alpha + {\cal O}(\alpha^2)
\label{eq:A:taylor:alpha}
\\
B(\alpha,\beta) &=& B_0(\beta) + B_1(\beta)\alpha + {\cal O}(\alpha^2)
\label{eq:B:taylor:alpha}
\eee
In other words:
\bee
P_0(\beta) &=& 2 A_0(\beta)
\label{eq:P0:beta:A0}
\\
P_1(\beta) &=&2A_1(\beta)-2B_0(\beta)
\label{eq:P1:beta:A1:B0}
\eee

\section{Calculating the bubble shape}
\label{sec:bubble:shape}

Functions $A_0(\beta)$ and $B_0(\beta)$
correspond to vanishing gravity ($\alpha=0$)
and can thus be determined by simply considering a spherical bubble
(Section \ref{sec:spherical:bubble:maintext}).

However, determining $A_1(\beta)$ (or equivalently $P_1(\beta)$)
requires to calculate the non-trivial bubble shape
in the presence of gravity
(Sections~\ref{sec:eq:bubble:shape}-\ref{sec:numeric:shape}).

\subsection{Spherical bubble}
\label{sec:spherical:bubble:maintext}

The spherical bubble case (zero gravity, $\alpha=0$)
is treated geometrically in Appendix~\ref{sec:spherical:cap}
and provides functions $A_0(\beta)$ and $B_0(\beta)$:
\bee
A_0(\beta) &=& \frac{ (U+4)^{2/3} - (U-4)^{2/3} } {U}
\label{eq:A0:U:beta}
\\
&&= 1 +{\cal O}(\beta^4)
\\
&& {\rm where}\,\, U = \sqrt{16 + \beta^6}
\nonumber\\
B_0(\beta) &=& \frac{1+\sqrt{1 -\beta^2 A_0^2}}{2A_0}
\\
&&= 1 -\frac14\beta^2 +{\cal O}(\beta^4)
\eee
The quantity $2A_0(\beta)$ is equal
to the zero-gravity component $P_0$ of the pressure,
as expressed by Eq.~\eqref{eq:P0:beta:A0}.
It is plotted as circles on Fig.~\ref{fig:bubble:a:shape:b:pressure}b.

Note that when the tube outlet radius goes to zero,
the above expressions go to unity: $A_0(0)=B_0(0)=1$.

\subsection{Equation for the bubble shape}
\label{sec:eq:bubble:shape}

In order to determine $A_1(\beta)$ (or $P_1(\beta)$),
let us generalize Eq.~\eqref{eq:p:dimension:apex} as:
\be
\pgas = \pliqout
+ \gamma\,C(s) + \rho g (\zout - z(s))
\label{eq:p:dimension:s}
\ee
where $C$ is the total curvature and $z$ the altitude at point $s$,
where $s$ is for instance the curvilinear distance from the top of the bubble.

Let $r(s)$ be the distance from the bubble axis
and $\psi(s)$ the angle between the tangent
to the bubble contour at point $s$ and the horizontal
(with the convention that $\psi(s)$ is positive).
The total curvature of such a axisymmetric shape can be shown to be:
\be
C(s) = \frac{{\rm d}\psi}{{\rm d}s} + \frac{\sin\psi(s)}{r(s)}
\label{eq:Ctot:dpsi:sinpsi:over:r}
\ee
The evolution of $r$ and $z$ along the contour
are trivially related to $\psi$:
\bee
\frac{{\rm d}r}{{\rm d}s} &=& \cos\psi
\label{eq:ode:r:cos:psi}
\\
\frac{{\rm d}z}{{\rm d}s} &=& -\sin\psi
\eee
The evolution of $\psi$ results
from Eqs.~(\ref{eq:p:dimension:s},\ref{eq:Ctot:dpsi:sinpsi:over:r}):
\bee
\frac{{\rm d}\psi}{{\rm d}s}
&=& \frac{2Q}{R} - \frac{\sin\psi}{r}
+ \frac{\alpha}{R^2}(z-\zapex)
\label{eq:evol:psi:Q}
\eee
where the constant $Q$ is defined by:
\bee
\frac{2Q}{R} &=& 
\frac{\pgas - \pliqout}{\gamma}
+ \frac{\alpha}{R^2}(\zapex-\zout)
\label{eq:Q:pgas:zout}
\eee
The evolution of the volume $V(s)$
of the bubble above altitude $z(s)$ is simply:
\be
\frac{{\rm d}V}{{\rm d}s}
= \pi r^2 \left|\frac{{\rm d}z}{{\rm d}s}\right|
= \pi r^2 \sin\psi
\ee
The boundary conditions at the apex ($s=0$)
and at the outlet ($s=\sout$) are:
\bee
r(0) &=& 0
\\
z(0) &=& \zapex = 0
\\
\psi(0) &=& 0
\\
V(0) &=& 0
\\
r(\sout) &=& \rout = \beta R
\label{eq:bc:rout}
\\
z(\sout) &=& \zout
\label{eq:bc:zout}
\\
V(\sout) &=& \frac{4\pi}{3} R^3
\label{eq:bc:Vout}
\eee
where $\zapex=0$ by convention
and where $\rout$ is related to $\beta$ through Eq.~\eqref{eq:def:beta:rout:R}.

\subsection{Non-dimensional bubble shape}

Note that in Eq.~\eqref{eq:evol:psi:Q},
$Q$ is unknown since it contains $\pgas-\pliqout$ and $\zout-\zapex$,
see Eq.~\eqref{eq:Q:pgas:zout}.
The curvilinear position $\sout$ of the outlet is also unknown,
and only for the correct value of $Q$
will boundary conditions~\eqref{eq:bc:rout} and~\eqref{eq:bc:Vout}
be satisfied for the same value of $\sout$.
Thus, for every value of $\alpha$ and $\beta$,
the system of Eqs.~(\ref{eq:ode:r:cos:psi}--\ref{eq:bc:Vout})
needs to be integrated a number of times with different values of $Q$
to obtain the correct $Q$ and hence a correct bubble shape and gas pressure.

In order to avoid these complications,
let us renormalize all distances with $R/Q$
(even though $Q$ is yet unknown):
\bee
\rr &=& r\,Q/R
\label{eq:def:rr}
\\
\rz &=& (z-\zapex)\,Q/R
\label{eq:def:rz}
\\
\rs &=& s\,Q/R
\\
\rV &=& V\,Q^3/R^3
\label{eq:def:rV}
\\
\ralpha &=& \alpha/Q^2
\label{eq:def:ralpha}
\eee
In terms of these new variables,
the system of differential equations reads:
\bee
\frac{{\rm d}\rr}{{\rm d}\rs} &=& \cos\psi
\label{eq:rode:r:cos:psi}
\\
\frac{{\rm d}\rz}{{\rm d}\rs} &=& -\sin\psi
\\
\frac{{\rm d}\psi}{{\rm d}\rs}
&=& 2 - \frac{\sin\psi}{\rr}
+ \ralpha\,\rz
\label{eq:rode:psi}
\\
\frac{{\rm d}\rV}{{\rm d}\rs}
&=& \pi \rr^2 \sin\psi
\label{eq:rode:V:r:psi}
\eee
For any given $\ralpha$, the above system
can be solved starting from the initial conditions:
\bee
\rr(0) &=& 0
\\
\rz(0) &=& 0
\\
\psi(0) &=& 0
\\
\rV(0) &=& 0
\label{eq:rode:ini:V}
\eee
The solution is obtained in the form of functions
$\rr(\ralpha,\rs)$,
$\rz(\ralpha,\rs)$,
$\psi(\ralpha,\rs)$,
$\rV(\ralpha,\rs)$.

\subsection{Outlet position and gas pressure}
\label{sec:outlet:position:gas:pressure}

Let us now define:
\be
\rbeta(\ralpha,\rs)
\equiv \frac{\rr(\ralpha,\rs)}
{\left(\frac{3}{4\pi}\rV(\ralpha,\rs)\right)^{1/3}}
\ee
Using this new function $\rbeta$
and definitions~\eqref{eq:def:rr} and~\eqref{eq:def:rV},
the position $\rsout(\ralpha,\beta)$ of the outlet
is obtained very simply as the value of $\rs$ where:
\be
\rbeta(\ralpha,\rs)
\equiv \frac{\rout}
{\left(\frac{3}{4\pi}V\right)^{1/3}}
= \beta
\label{eq:rbeta:ralpha:rs:beta}
\ee
Once $\rsout(\ralpha,\beta)$ is thus determined,
we define:
\bee
\rrout(\ralpha,\beta) &=& \rr(\ralpha,\rsout(\ralpha,\beta))
\label{eq:rrout:ralpha}
\\
\rzout(\ralpha,\beta) &=& \rz(\ralpha,\rsout(\ralpha,\beta))
\label{eq:rzout:ralpha}
\eee
And we obtain:
\bee
Q(\ralpha,\beta) &=& \rrout(\ralpha,\beta) \, R / \rout
\nonumber\\
&=& \rrout(\ralpha,\beta) / \beta
\label{eq:Q:ralpha:beta}
\eee
Using Eqs.~(\ref{eq:def:nondim:P},\ref{eq:Q:pgas:zout},%
\ref{eq:def:rz},\ref{eq:def:ralpha}),
the pressure $P$ and the gravity parameter $\alpha$
can be expressed from the results
of Eqs.~(\ref{eq:rrout:ralpha},\ref{eq:rzout:ralpha},\ref{eq:Q:ralpha:beta})
in terms of parameter $\ralpha$:
\bee
P(\ralpha,\beta) &=& \frac{R}{\gamma} \, (\pgas - \pliqout)
\nonumber\\
&=& (2 + \ralpha \, \rzout(\ralpha,\beta)) \, \rrout(\ralpha,\beta) / \beta
\label{eq:P:from:ralpha}
\\
\alpha(\ralpha,\beta)
&=& \ralpha \, Q^2(\ralpha,\beta)
= \ralpha \, \rrout^2(\ralpha,\beta) / \beta^2
\label{eq:alpha:from:ralpha}
\qquad
\eee

Expressions~(\ref{eq:P:from:ralpha},\ref{eq:alpha:from:ralpha})
are valid for all $\alpha$ values
within some range defined by Eq.~\eqref{eq:alpha:stability:alphamax}
where the bubble remains attached to the tube outlet.

Let us now first show the results
of an analytic derivation of $P_0(\beta)$ and $P_1(\beta)$
(Section~\ref{sec:analytic:shape})
and of its numeric counterpart (Section~\ref{sec:numeric:shape}).

\subsection{Analytic (near-spherical) shape}
\label{sec:analytic:shape}

Let us now decompose the functions
that appear in the system of
Eqs.~(\ref{eq:rode:r:cos:psi}-\ref{eq:rode:ini:V})
into the trivial solution when $\ralpha=0$
and a term that depends on $\ralpha$:
\bee
\rr(\ralpha,\rs) &=& \sin\rs + \ralpha \, \rr_1(\ralpha,\rs)
\label{eq:def:rr1}
\\
\rz(\ralpha,\rs) &=& \cos\rs -1 + \ralpha \, \rz_1(\ralpha,\rs)
\label{eq:def:rz1}
\\
\psi(\ralpha,\rs) &=& \rs + \ralpha \, \psi_1(\ralpha,\rs)
\label{eq:def:psi1}
\\
\rV(\ralpha,\rs) &=& \frac{\pi}{3}\left(
2-3\cos\rs +\cos^3\rs \right)
\nonumber\\
&&+ \ralpha \, \rV_1(\ralpha,\rs)
\label{eq:def:rV1}
\eee
where the initial conditions imply:
\be
\rr_1(0) = \rz_1(0) = \psi_1(0) = \rV_1(0) = 0
\ee

It is shown in Appendix~\ref{sec:appendix:analytic:shape}
that:
\bee
\rr_1 &=&
\frac13 \sin\rs \cos\rs
+\frac16 \sin\rs
-\frac12 \rs \cos\rs
\qquad
\label{eq:rr1:rs:maintext}
\\
\rz_1 &=& -\frac13\sin^2\rs
+\frac12\rs\sin\rs
\nonumber\\
&&+\frac23 \log\cos\frac{\rs}{2}
-\frac13\sin^2\frac{\rs}{2}
\label{eq:rz1:rs:maintext}
\eee
The dashed contours on Fig.~\ref{fig:bubble:a:shape:b:pressure}a
correspond to the dimensional version
of Eqs.~(\ref{eq:rr1:rs:maintext},\ref{eq:rz1:rs:maintext})
obtained through the non-dimensionalizing factor $Q$
provided by Eq.~\eqref{eq:Q:rO}
and plotted parametrically as a function of $\beta$
using Eq.~\eqref{eq:beta:rr:rO:ralpha:rdout}
to express it in terms of the same parameter $\rdout$.

Similarly, concerning the pressure $P$
defined by Eqs.~(\ref{eq:def:nondim:P},\ref{eq:P:def:taylor:P0:P1}),
as shown in Appendix~\ref{sec:appendix:pressure:and:derivatives},
explicit expressions for both the zero gravity limit $P_0$
and the first derivative $P_1$ are provided respectively
by Eqs.~\eqref{eq:P0:Om0:1:3} and~\eqref{eq:dPdalpha:beta}.
Using Eq.~\eqref{eq:beta:rr:rO:ralpha:rdout} again,
$P_0$ and $P_1$ can be plotted,
respectively as the solid and the dashed curves
on Fig.~\ref{fig:bubble:a:shape:b:pressure}b.

The limits $P_0$ and $P_1$ can be obtained easily,
as shown in Appendix~\ref{sec:appendix:small:outlet:limit}:
\bee
P_0(0) &=& 2
\\
P_1(0) &=& -\frac{4}{3}
\eee
These two values can be read out
on Fig.~\ref{fig:bubble:a:shape:b:pressure}b
as the value reached by both curves
when they meet the vertical axis $\beta=0$.
They are used in the approximate expression
announced as Eq.~\eqref{eq:pressure:2:frac43}.

A more elaborate expansion of the same expressions
is presented in Appendix~\ref{sec:taylor:expansion}
and yields Eq.~\eqref{eq:P:alpha:beta:result}
which can be expressed as:
\bee
P_0(\beta) &=& 2 + {\cal O}(\beta^4)
\label{eq:P0:beta:with:expansion}
\\
P_1(\beta) &=& -\frac43 + {\cal O}(\beta^2)
\label{eq:P1:beta:with:expansion}
\eee

\subsection{Numeric bubble shape}
\label{sec:numeric:shape}

As a complement to the analytic approach
of Section~\ref{sec:analytic:shape},
one can integrate numerically
Eqs.~\eqref{eq:rode:r:cos:psi} to~\eqref{eq:rode:V:r:psi}.

Because of the structure of Eq.~\eqref{eq:rode:psi}
which contains the ratio of $\sin\psi$ and $\rr$,
both going to zero at $\rs=0$,
we start with the following initial conditions:
\bee
\rr &=& \rs_1
\\
\rz &=& -\frac12 \, \rs_1^2
\\
\psi &=& \rs_1
\\
\rV &=& \frac{\pi}{4} \, \rs_1^4
\eee
We integrate using the explicit Runge–Kutta method of order (4,5),
more precisely GNU Octave's~\cite{gnuoctave2015} {\em ode45} function,
with a maximum integration step taken as equal to $\rs_1$.
We stop integration at the outlet position defined by $\beta$
as stated in Section~\ref{sec:outlet:position:gas:pressure},
then read $Q$, $P$ and $\alpha$
as prescribed by Eqs.~\eqref{eq:Q:ralpha:beta},
\eqref{eq:P:from:ralpha} and~\eqref{eq:alpha:from:ralpha} respectively.

Each integration is performed for a given triplet $(\ralpha,\beta,\rs_1)$.
For every pair of values $(\ralpha,\beta)$,
three integrations have been performed,
with $\rs_1$ equal to $10^{-3}$, $10^{-4}$ and $3.10^{-5}$.
The values $P(\ralpha,\beta)$ and $\alpha(\ralpha,\beta)$
have then been extrapolated to the limit $\rs_1\rightarrow 0$.
For each value of $\beta$, three such processes have been performed
with $\ralpha$ equal to $5.10^{-4}$, $2.10^{-4}$ and $10^{-4}$.
The resulting values of $P(\ralpha,\beta)$ and $\alpha(\ralpha,\beta)$
have been used to extrapolate $P(\alpha,\beta)$
and $\partial P(\alpha,\beta)/\partial\alpha$
to the limit $\alpha\rightarrow 0$,
so as to obtain $P_0(\beta)$ and $P_1(\beta)$.
This whole process has been carried out
for $\beta$ equal to $0.2$, $0.1$ and $0.05$
and the corresponding values of $P_0$ and $P_1$
are plotted on Fig.~\ref{fig:bubble:a:shape:b:pressure}b
as large circles and diamonds respectively (purple color).
Finally, values for $P_0(0)$ and $P_1(0)$
are shown in black color.
They were extrapolated from the corresponding values
for the three non-zero values of $\beta$.
The values thus obtained confirm the values $P_0(0)=2$ and $P_1(0)=-4/3$
adopted for the approximate expression announced
in Eq.~\eqref{eq:pressure:2:frac43}.

\appendix

\section{Truncated sphere}
\label{sec:spherical:cap}

In this Appendix, we consider the situation
with zero gravity ($\alpha=0$), hence with a purely spherical drop,
and calculate the drop radius of curvature and apex altitude
as a function of the outlet radius $\rout$.
The result is expressed in the form of $A=A_0(\beta)$ and $B=B_0(\beta)$
defined by Eqs.~(\ref{eq:apex:r:f1},\ref{eq:apex:z:f2},%
\ref{eq:A:taylor:alpha},\ref{eq:B:taylor:alpha}) with $\alpha=0$,
where $\beta$ is defined by Eq.~\eqref{eq:def:beta:rout:R}.

The bubble, whose radius is $R$ when purely spherical,
becomes a truncated sphere when attached to an outlet of radius $\rout$.
Let $\Rapex$ be the radius of the truncated sphere.

The height of the truncated part is:
\bee
H &=& 2\Rapex -(\zapex-\zout)
\label{eq:H:Rapex:zapex}
\eee
where $\zapex$ (resp. $\zout$)
is the altitude of the bubble apex (resp. tube outlet),
see Fig.~\ref{fig:bubble:a:shape:b:pressure}.
Pythagore:
\bee
\Rapex^2 &=& \rout^2 + (\Rapex-H)^2
\\
H &=& \Rapex - \sqrt{\Rapex^2-\rout^2}
\label{eq:H:Rapex:rout}
\eee

Using Eqs.~\eqref{eq:def:beta:rout:R} and~\eqref{eq:apex:r:f1}
to reformulate Eq.~\eqref{eq:H:Rapex:rout}:
\bee
\frac{H}{\Rapex} &=& 1 - \sqrt{1 - A_0^2\beta^2}
\label{eq:H:over:Rapex}
\eee

The volume of the truncated part
is that of a spherical cap
of height $H$ and radius of curvature $\Rapex$:
\be
\frac{\pi}{3}\,H^2\,(3\Rapex - H)
\ee

The condition that the initial drop of radius $R$
has the same volume as the truncated sphere of radius $\Rapex$
can be expressed as:
\be
\frac{4\pi}{3} R^3
= \frac{4\pi}{3} \Rapex^3
- \frac{\pi}{3}\,H^2\,(3\Rapex - H)
\label{eq:vol:ini:trunc:v0}
\ee

Using Eqs.~(\ref{eq:apex:r:f1},\ref{eq:H:over:Rapex})
and noting $Z=\sqrt{1-A_0^2\beta^2}$,
Eq.~\eqref{eq:vol:ini:trunc:v0} can be transformed as follows:
\bee
4(1-A_0^3)&=&(1-Z)^2(2+Z)
\\
4(1-A_0^3)&=&2-(2+A_0^2\beta^2)Z
\\
4A_0^3-2&=&(2+A_0^2\beta^2)\sqrt{1-A_0^2\beta^2}
\\
(4A_0^3-2)^2&=&4-3A_0^4\beta^4-A_0^6\beta^6
\eee
and finally, after dividing by $A_0^3$:
\be
(16+\beta^6)\,A_0^3 +3\beta^4\,A_0 -16 = 0
\label{eq:A:beta:polym3}
\ee
Defining:
\be
U = \sqrt{16 + \beta^6},
\ee
the solution to the third order
polynomial equation~\eqref{eq:A:beta:polym3} is:
\bee
A_0(\beta) &=& \frac{ (U+4)^{2/3} - (U-4)^{2/3} } {U}
\\
&=& 1 - \frac{1}{16}\beta^4 -\frac{1}{48}\beta^6 +o(\beta^6)
\label{eq:A:taylor:beta}
\eee

Using Eqs.~(\ref{eq:apex:r:f1},\ref{eq:apex:z:f2},%
\ref{eq:H:Rapex:zapex},\ref{eq:H:over:Rapex},\ref{eq:A:taylor:beta}):
\bee
B_0(\beta) &=& \frac{1+\sqrt{1 -\beta^2 A_0^2}}{2A_0}
\\
&=& 1 - \frac{1}{4}\beta^2 +\frac{1}{192}\beta^6 +o(\beta^6)
\eee

\section{Analytic (near-spherical) shape}
\label{sec:appendix:analytic:shape}

In the present Appendix, we derive the results
presented in Section~\ref{sec:analytic:shape}.

\subsection{First order functions}
\label{sec:analytic:rr1:rz1:psi1:rV1}

Using Eq.~\eqref{eq:def:psi1}, $\sin\psi$ and $\cos\psi$
can be expressed to first order in $\ralpha$:
\bee
\sin\psi = \sin\rs + \ralpha \, \psi_1\cos\rs + {\cal O}(\ralpha^2)
\label{eq:sin:psi:psi1}
\\
\cos\psi = \cos\rs - \ralpha \, \psi_1\sin\rs + {\cal O}(\ralpha^2)
\label{eq:cos:psi:psi1}
\eee
Inserting Eqs.~(\ref{eq:def:rr1}--\ref{eq:def:rV1})
and Eqs.~(\ref{eq:sin:psi:psi1},\ref{eq:cos:psi:psi1})
into Eqs.~(\ref{eq:rode:r:cos:psi}--\ref{eq:rode:V:r:psi}):
\bee
\frac{{\rm d}\rr_1}{{\rm d}\rs} &=& -\psi_1 \sin\rs
+ {\cal O}(\ralpha)
\label{eq:rode:r1:sins:psi1}
\\
\frac{{\rm d}\rz_1}{{\rm d}\rs} &=& -\psi_1 \cos\rs
+ {\cal O}(\ralpha)
\label{eq:rode:z1:coss:psi1}
\\
\sin\rs\,\frac{{\rm d}\psi_1}{{\rm d}\rs}
&=& \rr_1 -\psi_1\cos\rs \nonumber\\
&&-\sin\rs +\sin\rs\cos\rs
+ {\cal O}(\ralpha) \qquad
\label{eq:rode:psi1}
\\
\frac{{\rm d}\rV_1}{{\rm d}\rs}
&=& \pi (\psi_1 \cos\rs +2 \rr_1) \sin^2\rs
+ {\cal O}(\ralpha) \qquad
\label{eq:rode:V1:s:psi1:r1}
\eee
Let us differentiate Eq.~\eqref{eq:rode:psi1}
and combine it with Eq.~\eqref{eq:rode:r1:sins:psi1}:
\bee
&&\sin\rs\,\frac{{\rm d^2}\psi_1}{{\rm d}\rs^2}
+2\cos\rs\,\frac{{\rm d}\psi_1}{{\rm d}\rs}
\nonumber\\
&&\qquad\qquad
=-\cos\rs
+2\cos^2\rs
-1
\qquad\qquad
\eee
Multiplyling by $\sin\rs$:
\bee
&&\frac{{\rm d}}{{\rm d}\rs}\left(
\sin^2\rs\,\frac{{\rm d}\psi_1}{{\rm d}\rs}\right)
\nonumber\\
&&\qquad
=-\cos\rs\sin\rs
+2\cos^2\rs\sin\rs
-\sin\rs
\qquad
\eee
Integrating with respect to $\rs$:
\bee
\sin^2\rs\,\frac{{\rm d}\psi_1}{{\rm d}\rs}
&=&\frac12\cos^2\rs
-\frac23\cos^3\rs
+\cos\rs
-\frac{5}{6}
\qquad
\nonumber\\
&=&\left[\frac23\cos\rs-\frac12\right]\sin^2\rs
-\frac23\sin^2\frac{\rs}{2}
\eee
where the integration constant
was chosen to obtain zero when $\rs=0$.
Dividing by $\sin^2\rs$:
\bee
\frac{{\rm d}\psi_1}{{\rm d}\rs}
&=&\frac23\cos\rs-\frac12
-\frac{1}{6\cos^2\frac{\rs}{2}}
\eee
By integration:
\bee
\psi_1 &=& \frac23\sin\rs
- \frac{\rs}{2}
- \frac13 \tan\frac{\rs}{2}
\label{eq:psi1:rs}
\eee
Multiplying Eq.~\eqref{eq:psi1:rs} by $\sin\rs$
and integrating as suggested by Eq.~\eqref{eq:rode:r1:sins:psi1}
with the condition $\rr_1(0)=0$, we obtain:
\bee
\rr_1 &=&
\frac13 \sin\rs \cos\rs
+\frac16 \sin\rs
-\frac12 \rs \cos\rs
\qquad
\label{eq:rr1:rs}
\eee
Similarly, multiplying Eq.~\eqref{eq:psi1:rs} by $-\cos\rs$
or $1-2\cos^2\frac{\rs}{2}$,
as suggested by Eq.~\eqref{eq:rode:z1:coss:psi1}, we obtain:
\bee
-\psi_1 \cos\rs &=& -\frac23\sin\rs\cos\rs
+ \frac12\rs\cos\rs
\nonumber\\
&&- \frac13 \tan\frac{\rs}{2}
+ \frac23 \sin\frac{\rs}{2}\cos\frac{\rs}{2}
\\
&=&-\frac13(\sin^2\rs)^\prime
+\frac12(\rs\sin\rs+\cos\rs)^\prime
\qquad\nonumber\\
&&+\frac23(\log\cos\frac{\rs}{2})^\prime
+\frac23(\sin^2\frac{\rs}{2})^\prime
\label{eq:moins:psi1:cos:rs}
\eee
Injecting Eq.~\eqref{eq:moins:psi1:cos:rs}
into Eq.~\eqref{eq:rode:z1:coss:psi1}
and integrating with respect to $\rs$
while imposing that $\rz_1=0$ when $\rs=0$,
we obtain:
\bee
\rz_1 &=& -\frac13\sin^2\rs
+\frac12\rs\sin\rs
+\frac12(\cos\rs-1)
\nonumber\\
&&+\frac23 \log\cos\frac{\rs}{2}
+\frac23\sin^2\frac{\rs}{2}
\\
\rz_1 &=& -\frac13\sin^2\rs
+\frac12\rs\sin\rs
\nonumber\\
&&+\frac23 \log\cos\frac{\rs}{2}
-\frac13\sin^2\frac{\rs}{2}
\label{eq:rz1:rs}
\eee

Inserting Eqs.~(\ref{eq:psi1:rs},\ref{eq:rr1:rs})
into Eq.~\eqref{eq:rode:V1:s:psi1:r1}, we obtain:
\bee
\frac{1}{\pi}\frac{{\rm d}\rV_1}{{\rm d}\rs}
&=& \left[
  \frac43\sin\rs\cos\rs
-\frac32\rs\cos\rs\right.
\nonumber\\
&&\left.
+\frac13\tan\frac{\rs}{2}\right]\sin^2\rs
+ {\cal O}(\ralpha) \qquad
\eee
Integrating with respect to $\rs$
while imposing that $\rV_1=0$ when $\rs=0$, we obtain:
\bee
\rV_1 &=&
\frac{4\pi}{3}\sin^2\frac{\rs}{2}
+\frac{\pi}{3}\sin^2\rs\sin^2\frac{\rs}{2}
\nonumber\\
&&-\frac{\pi}{2}\rs\sin^3\rs
-\frac{\pi}{12}\sin^2(2\rs)
\qquad
\label{eq:rV1:rs}
\eee

\subsection{Outlet position}
\label{sec:outlet:position}

Since the outlet position is close to the lower pole of the sphere,
let us define:
\be
\rs = \pi - \rd
\label{eq:rs:pi:rd}
\ee
Using Eqs.~(\ref{eq:def:rr1},\ref{eq:rr1:rs},\ref{eq:rs:pi:rd}),
$\rr$ can be expressed as:
\bee
\rr &=& \rr_0 + \ralpha \, \rr_1
+{\cal O}(\ralpha^2)
\label{eq:rr:rd:ralpha}
\qquad
\\
\rr_0 &=& \sin\rd
\label{eq:rr0:sin:rd}
\\
\rr_1 = \frac{\partial\rr}{\partial\ralpha}
&=& -\frac{1}{3} \sin\rd \cos\rd
+\frac{1}{6} \sin\rd
\nonumber\\
&&+\frac{\pi}{2} \cos\rd
-\frac{1}{2} \rd \cos\rd
\\
\rr_0^\prime = \frac{\partial\rr}{\partial\rd}
&=& \cos\rd
\label{eq:rr0:rd}
\eee
Using Eqs.~(\ref{eq:def:rz1},\ref{eq:rs:pi:rd}),
the leading order of $\rz$ is:
\bee
\rz &=& \rz_0 +{\cal O}(\ralpha)
= -1 -\cos\rd +{\cal O}(\ralpha)
\label{eq:rz:rd:ralpha}
\qquad
\eee
Using Eqs.~(\ref{eq:def:rV1},\ref{eq:rV1:rs},\ref{eq:rs:pi:rd}),
the volume
\be
\rO=\frac{3}{4\pi}\rV
\label{eq:def:rO:rV:3:4pi}
\ee
can be expressed as:
\bee
\rO &=& \rO_0 + \ralpha \, \rO_1
+{\cal O}(\ralpha^2)
\label{eq:rO:rd:ralpha}
\qquad
\\
\rO_0 &=& \frac12 +\frac34\cos\rd -\frac14\cos^3\rd
\label{eq:rO:rd}
\\
\rO_1 &=& \cos^2\frac{\rd}{2}
+\frac14 \sin^2\rd\cos^2\frac{\rd}{2}
-\frac{3\pi}{8}\sin^3\rd
\qquad\nonumber\\
&&+\frac{3}{8}\rd\sin^3\rd
-\frac{1}{16}\sin^2(2\rd)
\label{eq:r1:rd}
\\
\rO_0^\prime
&=& \frac{\partial\rO_0}{\partial\rd}
= -\frac34\sin\rd +\frac34\sin\rd\cos^2\rd
\label{eq:rO0:rd}
\eee

The position $\rdout$ of the tube outlet
is defined by Eq.~\eqref{eq:rbeta:ralpha:rs:beta}
and can be expressed using Eq.~\eqref{eq:def:rO:rV:3:4pi}:
\bee
\beta &=& \frac{\rr}{\rO^{1/3}}(\ralpha,\rdout)
\label{eq:beta:rr:rO:ralpha:rdout}
\eee
Eq.~\eqref{eq:beta:rr:rO:ralpha:rdout}
can be used with Eq.~\eqref{eq:Q:ralpha:beta}
to express the non-dimensionalization factor:
\bee
Q &=& \frac{\rrout}{\beta}
= \rO^{1/3}(\ralpha,\rdout)
\label{eq:Q:rO}
\eee
where $\rO$ is provided by
Eqs.~(\ref{eq:rO:rd:ralpha},\ref{eq:rO:rd},\ref{eq:r1:rd}).

\subsection{Pressure and derivative}
\label{sec:appendix:pressure:and:derivatives}

Using Eq.~\eqref{eq:beta:rr:rO:ralpha:rdout},
Eqs.~(\ref{eq:P:from:ralpha},\ref{eq:alpha:from:ralpha})
can be transformed into:
\bee
\alpha &=& \ralpha \, \rO^{2/3}(\ralpha,\rdout)
\\
P &=& (2+\ralpha\rz_0)\,\rO^{1/3}(\ralpha,\rdout)
\label{eq:P:ralpha:rz0:rO:ralpha:rdout}
\eee

In the zero gravity limit ($\ralpha\rightarrow 0$),
Eq.~\eqref{eq:P:ralpha:rz0:rO:ralpha:rdout} simplifies into:
\bee
P_0 &=& 2\rO_0^{1/3}(\rdout)
\label{eq:P0:Om0:1:3}
\eee
where $\rO_0$ is given by Eq.~\eqref{eq:rO:rd}.
Similarly, $\beta$ is then given by:
\bee
\beta_0 &=& \frac{\rr_0(\rdout)}{\rO_0^{1/3}(\rdout)}
\label{eq:beta0:rr0:rO0:rdout}
\eee
where $\rr_0$ is given by Eq.~\eqref{eq:rr0:sin:rd}.
Then, using Eqs.~\eqref{eq:P0:Om0:1:3} and~\eqref{eq:beta0:rr0:rO0:rdout},
the pressure $P_0$ can be plotted as a function of $\beta_0$
using $\rdout$ as a parameter, which yields the solid curve $P_0(\beta)$
on Fig.~\ref{fig:bubble:a:shape:b:pressure}b.

In order to obtain $2A_1(\beta)-2B_0(\beta)
= \frac{\partial P}{\partial\alpha}(\alpha=0,\beta)$,
let us write the differentials of $P(\ralpha,\rd)$,
$\beta(\ralpha,\rd)$ and $\alpha(\ralpha,\rd)$:
\bee
{\rm d}P
&=& \frac{\partial P}{\partial\ralpha} {\rm d}\ralpha
+\frac{\partial P}{\partial\rd} {\rm d}\rd
\label{eq:dP:dralpha:drd}
\\
{\rm d}\beta
&=& \frac{\partial \beta}{\partial\ralpha} {\rm d}\ralpha
+\frac{\partial \beta}{\partial\rd} {\rm d}\rd
\label{eq:dbeta:dralpha:drd}
\\
{\rm d}\alpha
&=& \frac{\partial \alpha}{\partial\ralpha} {\rm d}\ralpha
+\frac{\partial \alpha}{\partial\rd} {\rm d}\rd
\label{eq:dalpha:dralpha:drd}
\qquad
\eee
Solving the system of Eqs.~(\ref{eq:dbeta:dralpha:drd},\ref{eq:dalpha:dralpha:drd})
for ${\rm d}\ralpha$ and ${\rm d}\rd$
and injecting them into Eq.~\eqref{eq:dP:dralpha:drd}, one obtains:
\bee
{\rm d}P
&=& \frac{
\frac{\partial\beta}{\partial\ralpha} 
\frac{\partial P}{\partial\rd}
-\frac{\partial\beta}{\partial\rd} 
\frac{\partial P}{\partial\ralpha}
}{
\frac{\partial\beta}{\partial\ralpha} 
\frac{\partial \alpha}{\partial\rd}
-\frac{\partial\beta}{\partial\rd} 
\frac{\partial \alpha}{\partial\ralpha}
}
\,{\rm d}\alpha
\qquad
\nonumber\\
&+&  \frac{
\frac{\partial P}{\partial\ralpha} 
\frac{\partial \alpha}{\partial\rd}
-\frac{\partial P}{\partial\rd} 
\frac{\partial \alpha}{\partial\ralpha}
}{
\frac{\partial\beta}{\partial\ralpha} 
\frac{\partial \alpha}{\partial\rd}
-\frac{\partial\beta}{\partial\rd} 
\frac{\partial \alpha}{\partial\ralpha}
}
\,{\rm d}\beta
\eee
In particular:
\be
P_1(\beta) =
\left.\frac{\partial P}{\partial\alpha}\right|_{\alpha=0}
=\left.\frac{
\frac{\partial\beta}{\partial\ralpha} 
\frac{\partial P}{\partial\rd}
-\frac{\partial\beta}{\partial\rd} 
\frac{\partial P}{\partial\ralpha}
}{
\frac{\partial\beta}{\partial\ralpha} 
\frac{\partial \alpha}{\partial\rd}
-\frac{\partial\beta}{\partial\rd} 
\frac{\partial \alpha}{\partial\ralpha}
}\right|_{\ralpha=0}
\label{eq:P1:in:terms:of:partial:derivatives}
\ee

In order to express Eq.~\eqref{eq:P1:in:terms:of:partial:derivatives}
more explicitely, one needs to evaluate partial derivatives
of $P$, $\beta$ and $\alpha$ with respect to $\ralpha$ and $\rd$.
Here, primes denote derivatives with respect to $\rd$:
\bee
\frac{\partial P}{\partial\ralpha}(0,\rdout)
&=& \rz_0\rO_0^{1/3}+\frac23\frac{\rO_1}{\rO_0^{2/3}}
\\
\frac{\partial P}{\partial\rd}(0,\rdout)
&=& \frac{\partial P_0}{\partial\rd}
= \frac23\frac{\rO_0^\prime}{\rO_0^{2/3}}
\eee
\bee
\frac{\partial\beta}{\partial\ralpha}(0,\rdout)
&=& \frac{\rr_1}{\rO_0^{1/3}}
-\frac{\rr_0\rO_1}{3\rO_0^{4/3}}
\nonumber\\
&=& \left( \frac{\rr_1}{\rr_0}
-\frac{\rO_1}{3\rO_0} \right) \beta_0
\\
\frac{\partial\beta}{\partial\rd}(0,\rdout)
&=& \frac{\rr^\prime}{\rO^{1/3}}
-\frac{\rr\rO^\prime}{3\rO^{4/3}}
\nonumber\\
&=& \frac{\rr_0^\prime}{\rO_0^{1/3}}
-\frac{\rr_0\rO_0^\prime}{3\rO_0^{4/3}}
\eee
\bee
\frac{\partial\alpha}{\partial\ralpha}(0,\rdout)
&=& \rO^{2/3}(0,\rdout)
+ \frac23\ralpha\frac{\rO_1}{\rO^{1/3}}(0,\rdout)
\nonumber\\
&=& \rO_0^{2/3}
\\
\frac{\partial\ralpha}{\partial\rd}(0,\rdout)
&=& \left(\ralpha\frac{\partial\rO^{2/3}}{\partial\rd}\right)(0,\rdout)
=0
\eee

Using the above expressions,
Eq.~\eqref{eq:P1:in:terms:of:partial:derivatives}
becomes:
\bee
P_1(\beta)&=&
(6\rO_0\rr_1\rO_0^\prime
-4\rr_0\rO_0^\prime\rO_1 
\nonumber\\
&&
+3\rr_0\rz_0\rO_0\rO_0^\prime
-9\rO_0^2\rz_0\rr_0^\prime
-6\rO_0\rr_0^\prime\rO_1
)
\nonumber\\
&&\times\frac{1}{3\rO_0^{4/3}\,(\rr_0\rO_0^\prime-3\rO_0\rr_0^\prime)}
\label{eq:dPdalpha:beta}
\eee
Once expressions for $\rr_0^\prime(\rd)$ and $\rO_0^\prime(\rd)$,
given by Eqs.~\eqref{eq:rr0:rd} and~\eqref{eq:rO0:rd},
as well as those for $\rO_0$, $\rr_0$, $\rz_0$, $\rO_1$ and $\rr_1$,
have been substituted into Eq.~\eqref{eq:dPdalpha:beta},
it provides an explicit expression of $P_1$ in terms of $\rdout$.
In the same way as $P_0$, again using $\beta(\rdout)$
given by Eq.~\eqref{eq:beta0:rr0:rO0:rdout},
$P_1$ can then be plotted parametrically as a function of $\beta$,
as shown on Fig.~\ref{fig:bubble:a:shape:b:pressure}b (dashed curve).

\subsection{Small outlet radius limit}
\label{sec:appendix:small:outlet:limit}

Let us now take the limit of a small needle outlet
($\beta \rightarrow 0$).

The following functions,
provided in Section~\ref{sec:outlet:position},
can be evaluated at $\beta=0$:
\bee
\rr_0(0)=0 & \rr_1(0)=\frac{\pi}{2} & \rr_0^\prime(0)=1
\\
\rz_0(0)=-2&&
\\
\rO_0(0)=1 & \rO_1(0)=1 & \rO_0^\prime(0)=0
\eee
Using these values,
Eqs.~\eqref{eq:P0:Om0:1:3} and~\eqref{eq:dPdalpha:beta}
yield the values of $P_0$ and $P_1$
in the limit of a very small tube outlet:
\bee
P_0(0) &=& 2\rO_0^{1/3}(0) = 2
\\
P_1(0) &=& -\frac{4}{3}
\eee
These values are used as coefficients in Eq.~\eqref{eq:pressure:2:frac43}.

A proper expansion at small $\beta$
is provided below, in Appendix~\ref{sec:taylor:expansion}.


\subsection{Small outlet radius expansion}
\label{sec:taylor:expansion}

Let us now use the decompositions
expressed by Eqs.~(\ref{eq:def:rr1}--\ref{eq:def:rV1})
and inject them into Eqs.~(\ref{eq:rbeta:ralpha:rs:beta},%
\ref{eq:P:from:ralpha},\ref{eq:alpha:from:ralpha})
in order to obtain an expansion for $P(\alpha,\beta)$
to be compared with Eqs.~(\ref{eq:P0:beta:A0}--\ref{eq:P1:beta:A1:B0}).

Using Eqs.~(\ref{eq:rr1:rs},\ref{eq:rs:pi:rd}):
\bee
\rr_1 &\simeq& \frac{\pi}{2} -\frac23\rd -\frac{\pi}{4}\rd^2 +{\cal O}(\rd^3)
\\
\rr &\simeq& \rd -\frac16\rd^3
+\ralpha\,\left(\frac{\pi}{2}-\frac23\rd-\frac{\pi}{4}\rd^2\right)
\nonumber\\
&&+{\cal O}(\rd^5,\ralpha\rd^3,\ralpha^2)
\label{eq:rr:rd:ralpha}
\qquad
\eee
Using Eqs.~(\ref{eq:rV1:rs},\ref{eq:rs:pi:rd}):
\bee
\frac{3}{4\pi} \rV_1 &\simeq& \left(1-\frac14\rd^2+{\cal O}(\rd^3)\right)
\\
\frac{3}{4\pi} \rV &\simeq& \left(1+{\cal O}(\rd^4)\right)
+\ralpha\,\left(1+{\cal O}(\rd^2)\right)
\qquad
\\
\left(\frac{3}{4\pi} \rV\right)^{1/3}
&\simeq& 1+\frac13\ralpha
+{\cal O}(\rd^4,\ralpha\rd^2,\ralpha^2)
\label{eq:rV:rd:ralpha}
\qquad
\eee
The position of the outlet is defined
by Eq.~\eqref{eq:rbeta:ralpha:rs:beta},
which can be expressed using
Eqs.~(\ref{eq:rr:rd:ralpha},\ref{eq:rV:rd:ralpha}):
\bee
0&=&\rr(\ralpha,\rdout)
- \beta \, \left(\frac{3}{4\pi} \rV(\ralpha,\rdout)\right)^{1/3}
\\
0&=& \rdout -\frac16\rdout^3
+\ralpha\,\left(\frac{\pi}{2}-\frac23\rdout-\frac{\pi}{4}\rdout^2\right)
\nonumber\\
&&+{\cal O}(\rdout^5,\ralpha\rdout^3,\ralpha^2)
\nonumber\\
&&-\beta-\frac13\ralpha\beta
+{\cal O}(\rdout^4\beta,\ralpha\rdout^2\beta,\ralpha^2)
\\
0&=& \rdout -\frac16\rdout^3 -\beta
\nonumber\\
&&+\ralpha\left( \frac{\pi}{2} -\frac23\rdout
-\frac{\pi}{4}\rdout^2 -\frac13\beta \right)
\quad
\nonumber\\
&&+{\cal O}(\rdout^5,\beta\rdout^4,\ralpha\rdout^3,\ralpha\rdout^2\beta,\ralpha^2)
\nonumber\\
0&=& \rdout \left(1 -\frac16\rdout^2 -\frac23\ralpha
-\frac{\pi}{4}\ralpha\rdout \right)
\nonumber\\
&&-\left( \beta -\ralpha\frac{\pi}{2} +\frac13\ralpha\beta \right)
\nonumber\\
&&+{\cal O}(\rdout^5,\beta\rdout^4,\ralpha\rdout^3,\ralpha\rdout^2\beta,\ralpha^2)
\qquad
\eee
Multiplying by
\bee
&&1+\frac16\rdout^2 +\frac23\ralpha +\frac{\pi}{4}\ralpha\rdout
\nonumber\\
&&\qquad
+{\cal O}(\rdout^4,\beta\rdout^3,\ralpha\rdout^2,\ralpha\rdout\beta,\ralpha^2)
\qquad
\eee
and neglecting terms of order $\ralpha^2$, one obtains:
\bee
0 &=& \rdout\left( 1 -\frac29\ralpha\rdout^2 -\frac{\pi}{12}\ralpha\rdout^3 \right)
\nonumber\\
&&-\left( \beta -\frac{\pi}{2}\ralpha +\frac13\ralpha\beta
+\frac16\beta\rdout^2 -\frac{\pi}{12}\ralpha\rdout^2 \right.
\nonumber\\
&&\left. +\frac{1}{18}\ralpha\beta\rdout^2 +\frac23\ralpha\beta
+\frac{\pi}{4}\ralpha\beta\rdout \right)
\nonumber\\
&&+{\cal O}(\rdout^5,\beta\rdout^4,\beta^2\rdout^3,
\nonumber\\
&&\qquad\ralpha\rdout^3,\ralpha\beta\rdout^2,\ralpha\beta^2\rdout,\ralpha^2)
\eee
Hence:
\bee
\rdout &=& \beta +\frac16\beta\rdout^2
-\frac{\pi}{2}\ralpha +\ralpha\beta
\nonumber\\
&&+\frac{\pi}{4}\ralpha\beta\rdout
-\frac{\pi}{12}\ralpha\rdout^2 
\nonumber\\
&&+{\cal O}(\rdout^5,\beta\rdout^4,\beta^2\rdout^3,
\nonumber\\
&&\qquad\ralpha\rdout^3,\ralpha\beta\rdout^2,\ralpha\beta^2\rdout,\ralpha^2)
\eee
This shows that the dominant terms are
$\rdout\simeq\beta-\frac{\pi}{2}\ralpha$.
Hence, the terms containing $\rdout$ and the neglected terms
can be expressed in terms of $\beta$:
\bee
\rdout &=& \beta +\frac16\beta\left(\beta-\frac{\pi}{2}\ralpha\right)^2
-\frac{\pi}{2}\ralpha +\ralpha\beta
\nonumber\\
&&+\frac{\pi}{4}\ralpha\beta^2
-\frac{\pi}{12}\ralpha\beta^2 
\nonumber\\
&&+{\cal O}(\beta^5,\ralpha\beta^3,\ralpha^2)
\\
\rdout &=& \beta +\frac16\beta^3
-\frac{\pi}{2}\ralpha +\ralpha\beta
\nonumber\\
&&+{\cal O}(\beta^5,\ralpha\beta^3,\ralpha^2)
\label{eq:rdout:beta:ralpha}
\eee
Injecting Eq.~\eqref{eq:rdout:beta:ralpha}
into Eqs.~(\ref{eq:def:rz1},\ref{eq:rs:pi:rd},\ref{eq:rr:rd:ralpha}),
we obtain:
\bee
\ralpha\rzout &=& -\ralpha -\ralpha\cos\beta +{\cal O}(\ralpha^2)
\nonumber\\
&=& -2\ralpha +\frac12\ralpha\beta^2
+{\cal O}(\ralpha\beta^4,\ralpha^2)
\label{eq:ralpharzout:ralpha:beta}
\\
\rrout &=&
\left(\beta +\frac16\beta^3 -\frac{\pi}{2}\ralpha +\ralpha\beta \right)
\nonumber\\
&&-\frac16\left( \beta -\frac{\pi}{2}\ralpha \right)^3
+\ralpha\,\left(\frac{\pi}{2}-\frac23\beta-\frac{\pi}{4}\beta^2\right)
\nonumber\\
&&+{\cal O}(\beta^5,\ralpha\beta^3,\ralpha^2)
\nonumber\\
&=& \beta +0\beta^3 +\frac13\ralpha\beta +0\ralpha\beta^2
\nonumber\\
&&+{\cal O}(\beta^5,\ralpha\beta^3,\ralpha^2)
\\
\frac{\rrout}{\beta} &=& 1 +\frac13\ralpha 
+{\cal O}(\beta^4,\ralpha\beta^2,\ralpha^2)
\qquad
\label{eq:rr:ralpha:beta}
\eee

Injecting Eqs.~(\ref{eq:ralpharzout:ralpha:beta},\ref{eq:rr:ralpha:beta})
into Eqs.~(\ref{eq:P:from:ralpha},\ref{eq:alpha:from:ralpha}):
\bee
P(\ralpha,\beta)
&=& 2 -\frac43\ralpha 
+{\cal O}(\beta^4,\ralpha\beta^2,\ralpha^2)
\label{eq:P:ralpha:beta:result}
\\
\alpha(\ralpha,\beta)
&=& \ralpha +{\cal O}(\ralpha\beta^4,\ralpha^2)
\label{eq:alpha:ralpha:beta:result}
\eee
Substituting Eq.~\eqref{eq:alpha:ralpha:beta:result}
into Eq.~\eqref{eq:P:ralpha:beta:result}:
\bee
P(\alpha,\beta)
&=& 2 -\frac43\alpha 
+{\cal O}(\beta^4,\alpha\beta^2,\alpha^2)
\label{eq:P:alpha:beta:result}
\eee
In other words, $P_0(\beta)=P(0,\beta)$
and $P_1(\beta)=(\partial P/\partial\alpha)|_{(0,\beta)}$
are given by Eqs.~\eqref{eq:P0:beta:with:expansion}
and~\eqref{eq:P1:beta:with:expansion}, as announced.

\bibliography{2016_shape}
\bibliographystyle{plain}

\end{document}